\pdfoutput=1
\documentclass[runningheads]{llncs}
\usepackage{graphicx}

\usepackage{hyperref}

\usepackage{cite}
\usepackage{amsmath}   
\usepackage{amsfonts}    
\usepackage{enumerate}  
\usepackage{rotating}
\usepackage{comment}
\usepackage{pifont}

\usepackage{color,soul}
\usepackage{float}
\usepackage[table,xcdraw]{xcolor}
\usepackage{multirow}
\usepackage{hhline}
\usepackage{arydshln}
\usepackage{booktabs}
\usepackage{subcaption}
\usepackage{xspace}
\newcommand*{\eg}{\emph{e.g.}\@\xspace}
\newcommand*{\ie}{\emph{i.e.}\@\xspace}
\newcommand*{\proposed}{\texttt{AdLocUI}\@\xspace}
\newcommand*{\baseline}{Yeung et al.\cite{yeung2021learning}\@\xspace}

\begin{document}
\title{Adaptive 3D Localization of 2D Freehand Ultrasound Brain Images}
%
\titlerunning{Adaptive 3D Localization of 2D Freehand Ultrasound Brain Images}


\author{Pak-Hei Yeung\inst{1,2} \and
Moska Aliasi\inst{3} \and
Monique Haak\inst{3} \and
the INTERGROWTH-21$\textsuperscript{st}$ Consortium\inst{4} \and
Weidi Xie\inst{5,6} \and
Ana I.L. Namburete\inst{2}}

\authorrunning{PH. Yeung et al.}

\institute{Oxford Machine Learning in NeuroImaging Lab, Department of Computer Science, University of Oxford, Oxford, United Kingdom
\and
Department of Engineering Science, Institute of Biomedical Engineering, University of Oxford, Oxford, United Kingdom\\
\and
Division of Fetal Medicine, Department of Obstetrics, Leiden University Medical Center, 2333 ZA Leiden, The Netherlands
\and
Nuffield Department of Women's and Reproductive Health, University of Oxford, Oxford, United Kingdom
\and
Shanghai Jiao Tong University, Shanghai, China
\and
Visual Geometry Group, Department of Engineering Science, University of Oxford, Oxford, United Kingdom\\
\email{pak.yeung@pmb.ox.ac.uk}
}

\maketitle              
\begin{center}
\url{https://pakheiyeung.github.io/AdLocUI_wp/}
\end{center}

%

\begin{abstract}
Two-dimensional (2D) freehand ultrasound is the mainstay in prenatal care and fetal growth monitoring.
The task of matching corresponding cross-sectional planes in the 3D anatomy for a given 2D ultrasound brain scan is essential in freehand scanning, but challenging.
We propose \proposed,
a framework that \textbf{Ad}aptively \textbf{Loc}alizes 2D \textbf{U}ltrasound \textbf{I}mages in the 3D anatomical atlas  \emph{without} using any external tracking sensor.
We first train a convolutional neural network with 2D slices sampled from co-aligned 3D ultrasound volumes to predict their locations in the 3D anatomical atlas.
Next, we fine-tune it with 2D freehand ultrasound images using a novel \textbf{unsupervised cycle consistency},
which utilizes the fact that the overall displacement of a sequence of images in the 3D anatomical atlas is equal to the displacement from the first image to the last in that sequence.
We demonstrate that \proposed can adapt to three different ultrasound datasets, acquired with different machines and protocols,
and achieves significantly better localization accuracy than the baselines.
\proposed can be used for sensorless 2D freehand ultrasound guidance by the bedside.
The source code is available at \url{https://github.com/pakheiyeung/AdLocUI}.


\end{abstract}

\begin{figure*}
\centering
\includegraphics[width=\textwidth]{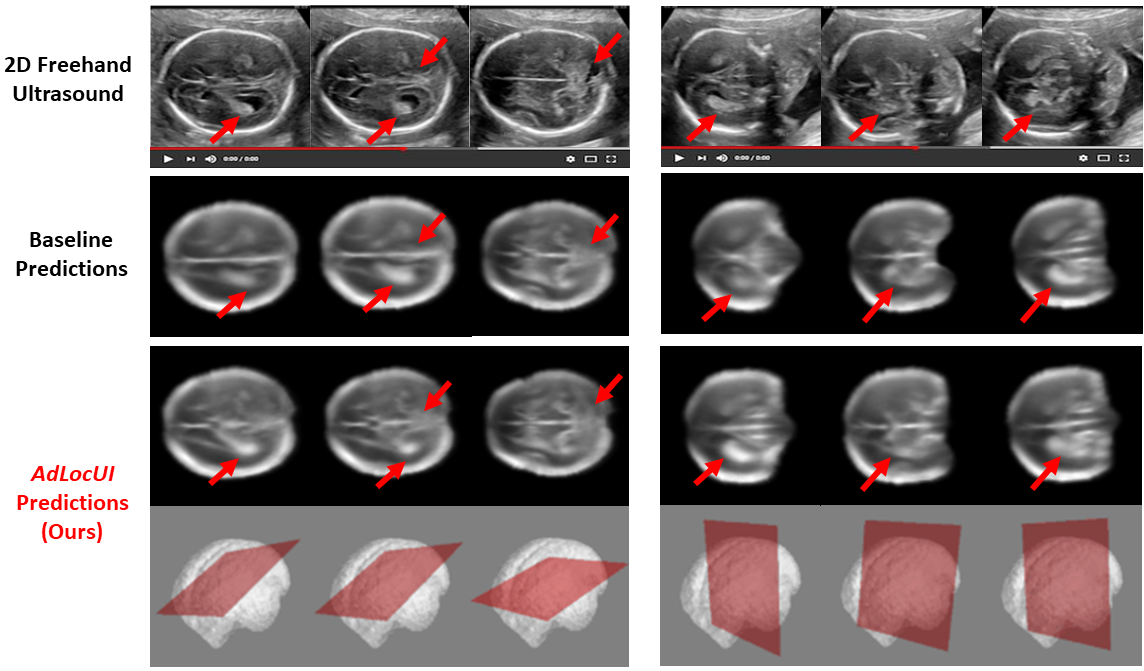}
\caption{
Localization of 2D freehand ultrasound images in the 3D anatomical atlas (\ie fetal brain).
2D slices sampled from the 3D atlas using image locations predicted by the baseline \cite{yeung2021learning} and \proposed are presented, where our predictions show better correspondence 
(\ie emphasized by the red arrows) with the ultrasound images,
suggesting more accurate 3D localization prediction by \proposed. 
} 
\label{fig:qualitative}
\end{figure*}

%
%
%
\section{Introduction}
\label{sec:introduction}
Two-dimensional (2D) freehand ultrasonography
is one of the most routinely deployed medical imaging modalities in prenatal care.
The nature of ultrasound images is unique when compared to other modalities.
While magnetic resonance imaging (MRI) and computerised tomography (CT)
capture the complete 3D anatomy,
each 2D ultrasound image is just a 2D cross-sectional view of an inherently 3D anatomy.
It also differs from other 2D imaging modalities,
such as X-ray which captures a projectional view of the 3D body.
Such fundamental differences make image localization in the 3D anatomy a unique but important task for 2D freehand ultrasonography, especially for neuroimaging \cite{Paladini2007} (Fig.~\ref{fig:qualitative}).

Experienced sonographers 
are often able to locate the 2D ultrasound images by mentally reconstructing the 3D anatomy~\cite{Goncalves2005}.
However, training a network to achieve this is very challenging,
mainly due to the difficulty of acquiring the training data 
(\ie 2D freehand ultrasound images and their corresponding locations in the 3D anatomy).
In this paper, 
we overcome this limitation by using only a small number (\ie 50) of 3D ultrasound volumes, 
co-aligned to a common 3D anatomical atlas,
for the training. 
It can then be fine-tuned by our proposed \textbf{unsupervised cycle consistency} to adapt to 2D ultrasound images acquired from different machines and protocols.
The \emph{only} manual annotation required in our work is the co-alignment of the 3D training volumes,
which could be further automated with volumetric registration algorithms, such as~\cite{namburete2018fully}. 

A related task,
namely standard plane detection,
has been attempted in numerous prior works,
using convolutional neural networks (ConvNet)~\cite{Baumgartner2017, Gao2017, gao2020label} 
and reinforcement learning~\cite{li2021autonomous, dou2019agent}.
A recent study~\cite{droste2020automatic} extended standard plane detection to a guidance system using an external motion sensor.
Despite their excellent performance, 
the overarching problem of locating a 2D freehand scan in the 3D anatomy remains unexplored.
Volumetric reconstruction with motion-tracked probe is another related research topic \cite{mohamed2019survey, mozaffari2017freehand}.
However, freehand ultrasound scanning of the fetus is challenged by the fact that the subject is not stationary, particularly before the third trimester. 
A tracking sensor, therefore, can only record the probe position but not the plane position due to the relative motion between the fetus and the probe. 
This limits the tracking sensors’ practical application
in our problem setting.
Inspired by \cite{hou20183},
Yeung et al.~\cite{yeung2021learning} proposed to use 2D slices sampled from 3D ultrasound volumes to train a network to predict the 3D locations of 2D ultrasound images.
Our work extends this
by proposing a framework that adapts the trained network to generalize to 2D ultrasound images acquired from diverse machines and protocols,
which are essential for realistic scenarios in which data are acquired from different clinical centres.

In this paper,
we propose \proposed,
a framework that \textbf{Ad}aptively \textbf{Loc}alizes 2D \textbf{U}ltrasound \textbf{I}mages
in a predefined 3D anatomical atlas (\ie fetal brain).
Our work makes the following contributions:
\emph{firstly},
we propose a framework for the aforementioned localization task. 
We demonstrate that a \emph{single} model, 
trained with minimal manual annotation (\ie co-alignment of a set of 3D volumes),
can be fine-tuned in an \textbf{unsupervised} manner and adapted to 3 different datasets of ultrasound images,
acquired from diverse machines and acquisition protocols differing from those of the training data.
\emph{Secondly},
we propose a novel way to fine-tune the trained model to adapt to the target domain 2D ultrasound images,
which utilizes the fact that the overall displacement of a sequence of images in the 3D anatomical atlas is equal to the displacement from the first image to the last in that sequence.
As our \emph{third} contribution,
we show, with ablation studies,
that the introduction of our proposed fine-tuning step leads to a significant improvement on localization accuracy when compared to the baseline~\cite{yeung2021learning},
and that fine-tuned by popular domain adaptation (DA) algorithms~\cite{sun2016deep, long2015learning, ganin2016domain}. 
Our framework can be used for \emph{sensorless} volumetric reconstruction \cite{yeung2021implicitvol} and freehand guidance for training and facilitating more objective analysis and diagnosis.




\begin{figure*}[t]
\centering
\includegraphics[width=\textwidth]{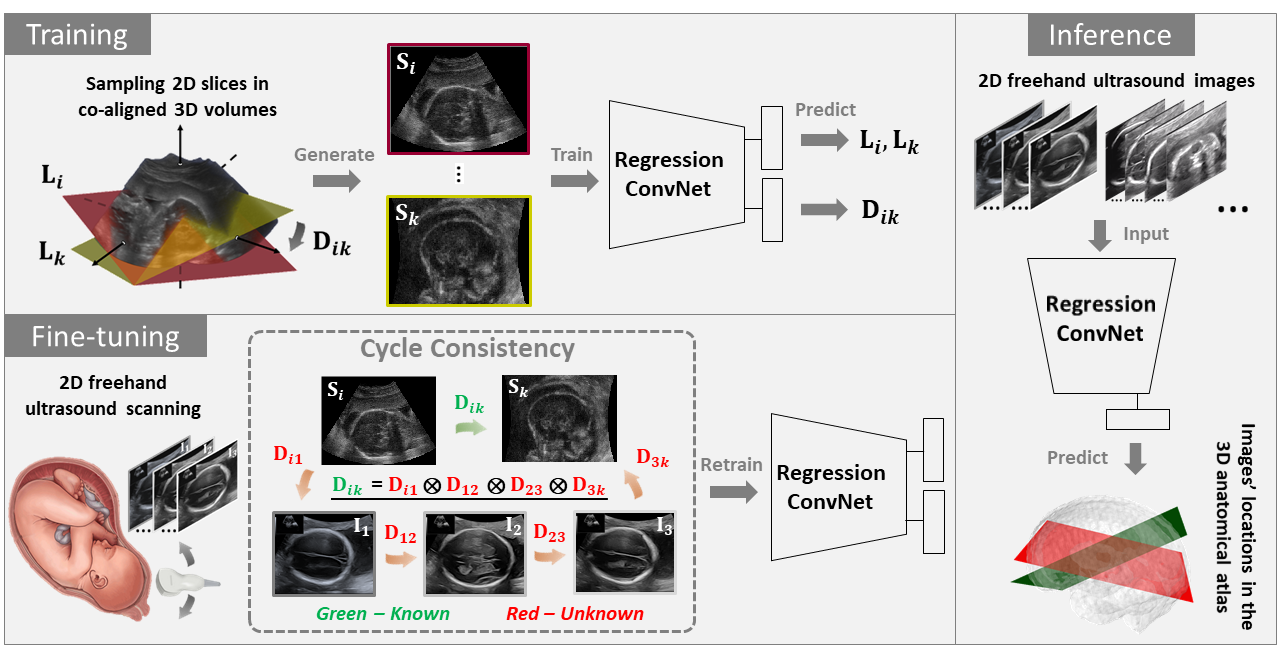}
\caption{Pipeline of our proposed framework, \proposed. 
During training, 
2D slices, ${\mathbf{S}_i}$,
sampled from co-aligned 3D volumes are used to train a regression ConvNet to predict the locations, ${\mathbf{L}_i}$, and displacement ${\mathbf{D}_{ik}}$, of the 2D slices in the 3D anatomical atlas.
The ConvNet is then fined-tuned \textbf{unsupervisedly} with 2D freehand ultrasound images, 
${\mathbf{I}_i}$, based on the proposed cycle consistency.
We can then use the fine-tuned ConvNet to localize ${\mathbf{I}_i}$ of the same domain (\ie acquired with the same machines and protocols) in the predefined 3D anatomical atlas. 
} 
\label{fig:pipeline}
\end{figure*}

\section{Methods}
\label{sec:method}

\subsection{Problem Setup}
\label{sec:setup}
We consider each ultrasound acquisition from different machines as a different domain.
In general, 
given a sequence or set of $m$ 2D ultrasound images,
$\mathcal{I} = \{\mathbf{I}_1, \mathbf{I}_2, \dots, \mathbf{I}_m\}$,
acquired from any domain,
our goal is to predict their locations,
$\mathcal{L}_{Img} = \{\mathbf{L}_1, \mathbf{L}_2, \dots, \mathbf{L}_m\}$,
in a predefined 3D anatomical atlas,
$\mathbb{R}^{3}_{atlas}$.

We formulate this problem in 3 stages (Fig. \ref{fig:pipeline}). 
In \emph{training},
we train a regression ConvNet,
$\psi(\cdot;\theta)$,
parametrized by $\theta$, 
with $n$ 2D slices,
$\mathcal{S} = \{\mathbf{S}_1, \mathbf{S}_2, \dots, \mathbf{S}_n\}$,
sampled from the corresponding plane locations,
$\mathcal{L}_S = \{\mathbf{L}_1, \mathbf{L}_2, \dots, \mathbf{L}_n\}$,
of a set of 3D ultrasound volumes 
co-aligned in $\mathbb{R}^{3}_{atlas}$.
After that,
we retrain (\ie \emph{fine-tune})
$\psi(\cdot;\theta)$ with $\mathcal{S}$ and 
$\mathcal{I}$,
using cycle consistency in an \textbf{unsupervised} manner.
$\psi(\cdot;\theta)$ can then be used on $\mathcal{I}$ or images of the same domain as $\mathcal{I}$ during \emph{inference}.
For clarification, 
we will refer to 2D slice sampled from the 3D training volumes as $\mathbf{S}$ and the target domain 2D ultrasound image as $\mathbf{I}$.
\subsection{Training with Sampled 2D Slices from 3D Volumes}
\label{sec:training}
Conventionally,
training $\psi(\cdot;\theta)$,
requires paired training data 
(\ie $\{{\mathbf{I}_i}, {\mathbf{L}_i}\}$),
where ${\mathbf{L}_i}$ (parameterization of ${\mathbf{L}}$ is detailed below) needs to be manually annotated,
which is very challenging and time-consuming. 
A prior study \cite{yeung2021learning}
proposed to use 
2D slices,
$\mathcal{S}$,
sampled from aligned 3D ultrasound volumes,
as the training data.
Therefore, the corresponding plane locations, $\mathcal{L}_S$,
of the 2D slices are automatically known,
voiding the need for further manual annotation.
We adopt the same strategy in this study.
\\
\\
\noindent \textbf{Data preparation pipeline.}
We affinely registered a set of 3D ultrasound volumes,
to a common predefined anatomical atlas,
$\mathbb{R}^{3}_{atlas}$,
either manually,
or by alignment algorithms such as~\cite{namburete2018fully},
followed by minor manual correction.
This is the \textbf{only} manual annotation required by \proposed.
2D slices, $\mathcal{S}$, were then randomly sampled from the aligned volumes,
using Fibonacci sphere sampling of polar coordinates \cite{hou20183},
on the fly during training.
The details are described in \cite{yeung2021learning}.
\\
\\
\noindent \textbf{Training objectives.}
With a set of $n$ paired training data,
$\{{\mathbf{S}_i, \mathbf{L}_i}\}_{i=1}^n$,
a regression ConvNet, $\psi(\cdot;\theta)$,
is trained. 
$\psi(\cdot;\theta)$ is composed of 3 parts,
namely the encoder $\psi_{enc}(\cdot;\theta_{enc})$, 
location prediction $\psi_{loc}(\cdot;\theta_{loc})$ and displacement prediction $\psi_{disp}(\cdot;\theta_{disp})$.
First, 
$\mathcal{S}$
are randomly augmented 
by 
scaling, in-plane translation, contrast adjustment 
and random noise. 
A feature vector, $\mathbf{v}_i$, 
is then generated by the encoder part,
$\psi_{enc}(\cdot;\theta_{enc})$, 
for each $\mathbf{S}_i$:
\begin{equation}
\label{eq:encoder}
[\mathbf{v}_1,\ \mathbf{v}_2,...,\ \mathbf{v}_n] = [\psi_{enc}(\mathbf{S}_1;\theta_{enc}),\ \psi_{enc}(\mathbf{S}_2;\theta_{enc}),\ ...,\ \psi_{enc}(\mathbf{S}_n;\theta_{enc})]
\end{equation}
Similar to \cite{yeung2021learning},
the feature vectors, 
$\{\mathbf{v}_1, \mathbf{v}_2, \dots, \mathbf{v}_n\}$,
are used to predict the plane locations,
$\mathcal{L}_S$,
by the location prediction part,
$\psi_{loc}(\cdot;\theta_{loc})$:
\begin{equation}
\label{eq:location}
[\hat{\mathbf{L}}_1,\ \hat{\mathbf{L}}_2,...,\ \hat{\mathbf{L}}_n] = [\psi_{loc}(\mathbf{v}_1;\theta_{loc}),\ \psi_{loc}(\mathbf{v}_2;\theta_{loc}),\ ...,\ \psi_{loc}(\mathbf{v}_n;\theta_{loc})]
\end{equation}
where $\hat{\ }$ indicates predicted values.
Unlike \cite{yeung2021learning},
we simultaneously predict $\mathbf{L}_i$ and the \emph{displacement},
$\mathbf{D}_{ik}$,
between each pair of slices, 
$\mathbf{S}_{i}$ and $\mathbf{S}_{k}$,
in $\mathbb{R}^{3}_{atlas}$
by the displacement prediction part,
$\psi_{disp}(\cdot;\theta_{disp})$:
\begin{equation}
\label{eq:relative}
[...,\ \hat{\mathbf{D}}_{ik},...,\ \hat{\mathbf{D}}_{nn}] = [...,\  \psi_{disp}(\mathbf{v}_i,\mathbf{v}_k;\theta_{disp}),\ ...,\ \psi_{disp}(\mathbf{v}_n,\mathbf{v}_n;\theta_{disp})]
\end{equation}
\\
\\
\noindent \textbf{Parameterization of $\mathbf{L}$ and $\mathbf{D}$.}
Following the practice of \cite{yeung2021learning},
we parameterize the plane location,  
$\mathbf{L}_i \in \mathcal{D}^{3 \times 3}$,
by three anchor points (\ie their \emph{x, y} and \emph{z} coordinates),
namely the \emph{top right}, \emph{top left} and \emph{bottom right} corners,
of $\mathbf{S}_i$.
The displacement,
$\mathbf{D}_{ik}$,
from
$\mathbf{S}_{i}$ to $\mathbf{S}_{k}$,
in $\mathbb{R}^{3}_{atlas}$
is therefore parameterized as $(\mathbf{L}_{i} - \mathbf{L}_{k})$.
There are other parameterization methods,
such as Euler angles and quaternions \cite{hou2017predicting, hou20183},
which may be investigated in future work.
\\
\\
\noindent \textbf{Training loss ($\mathit{l}_t$).}
We use weighted mean least-squared error (MSE) as the loss function for the multi-task learning:
\begin{equation}
\label{eq:train_loss}
\mathit{l}_t = w_{L}\cdot\text{MSE}\left(\hat{\mathbf{L}},\mathbf{L}\right) + w_{D}\cdot\text{MSE}\left(\hat{\mathbf{D}},\mathbf{D}\right)
\end{equation}
where $w_{L}$ and $w_{D}$ are the weights of the respective MSE loss.

\subsection{Fine-tuning with 2D Ultrasound Images}
\label{sec:finetuning}
The trained ConvNet, $\psi(\cdot;\theta)$,
can then be fine-tuned with a new set of $m$ 2D ultrasound images,
$\mathcal{I}$,
acquired from any domain. 
The retraining relies on \emph{cycle consistency}
and uses both the training data 
(\ie $\{{\mathcal{S}, \mathcal{L}_S}\}$) and 
the new set of images, $\mathcal{I}$,
\emph{without} further manual annotation.
\\
\\
\noindent \textbf{Cycle consistency.}
Although the plane locations, 
${\mathcal{L}_{Img}}$
of the new $\mathcal{I}$ are unknown,
we know, by cycle consistency, that the overall displacement, $\mathbf{D}$, 
of a sequence of images
in $\mathbb{R}^{3}_{atlas}$
must be equal to $\mathbf{D}$ from the first image to the last of that sequence.
For example,
as illustrated in Fig.~\ref{fig:pipeline},
the overall displacement (i) 
$\mathbf{S}_{i} \rightarrow \mathbf{I}_{1}$ ($\mathbf{D}_{i1}$), and
$\mathbf{I}_{1} \rightarrow \mathbf{I}_{2}$ ($\mathbf{D}_{12}$), and
$\mathbf{I}_{2} \rightarrow \mathbf{I}_{3}$ ($\mathbf{D}_{23}$), and
$\mathbf{I}_{3} \rightarrow \mathbf{S}_{k}$ ($\mathbf{D}_{3k}$)
is equal to (ii)
$\mathbf{S}_{i} \rightarrow \mathbf{S}_{k}$ ($\mathbf{D}_{ik}$).
While every $\mathbf{D}$ in (i) is unknown, 
$\mathbf{D}_{ik}$ in (ii) is known from the original training data.
Therefore, 
we can construct the cycle consistency loss ($\mathit{l}_c$) with this equality to retrain $\psi(\cdot;\theta)$:
%
\begin{equation}
\label{eq:cycle_loss}
\mathit{l}_c = \text{MSE}\left(\hat{\mathbf{D}}_{i1} \otimes \hat{\mathbf{D}}_{12} \otimes \hat{\mathbf{D}}_{23} \otimes \hat{\mathbf{D}}_{3k},\ 
\mathbf{D}_{ik}\right) 
\end{equation}
where $\otimes$ depends on the choice of the parameterization of $\mathbf{D}$
and, hence, $\otimes$ is simply \emph{subtraction} here
(similar to the derivation of $\mathbf{D}$ from $\mathbf{L}$ described in Section~\ref{sec:training}).
When predicting two consecutive displacements (\eg $\mathbf{D}_{12}$ and $\mathbf{D}_{23}$),
the common image involved (\ie $\mathbf{I}_{2}$) is augmented differently,
which coincides with the recent self-supervised and unsupervised learning studies \cite{chen2020simple,grill2020bootstrap,oord2018representation}
that emphasize the importance of data augmentation.
Our proposed unsupervised cycle consistency mechanism is also conceptually different from that proposed in other ultrasound imaging studies, 
such as \cite{kz2020semi,delaunay2020unsupervised}.
\\
\\
\noindent \textbf{Fine-tuning loss ($\mathit{l}_f$).}
Since the goal of \proposed is to predict the corresponding plane location,
$\mathcal{L}_{Img}$,
of $\mathcal{I}$,
relying solely on the cycle consistency loss, $\mathit{l}_c$, (\ie supervise only on ${\mathbf{D}}$) may diverge the prediction or even fall into trivial solutions \cite{zhou2016learning}.
Therefore, we add the original training loss, $\mathit{l}_t$ (Eq.~\ref{eq:train_loss}), to regulate the retraining
and the overall fine-tuning loss, $\mathit{l}_f$ is:
\begin{equation}
\label{eq:fine_loss}
\mathit{l}_f = w_{c}\cdot \mathit{l}_c + \mathit{l}_t
\end{equation}
where $w_c$ is the weight of the  cycle consistency loss, $\mathit{l}_c$.

\subsection{Inference}
\label{sec:inference}
The fine-tuned ConvNet,
$\psi(\cdot;\theta)$,
can be used on the set of 2D ultrasound images,
$\mathcal{I}$,
or other 2D ultrasound images of the same domain
(\ie acquired from the same machine) to predict their corresponding locations,
$\mathcal{L}_{Img}$,
in the predefined 3D anatomical atlas,
$\mathbb{R}^{3}_{atlas}$:
\begin{equation}
\label{eq:inference}
[\hat{\mathbf{L}}_1,\ \hat{\mathbf{L}}_2,...,\ \hat{\mathbf{L}}_m] = [\psi(\mathbf{I}_1;\theta),\ \psi(\mathbf{I}_2;\theta),\ ...,\ \psi(\mathbf{I}_m;\theta)]
\end{equation}
%



\section{Experimental Design}
\label{sec:experiment}
\proposed and other baseline approaches were first trained with 2D slices,
$\mathcal{S}$, sampled from 50 3D volumes acquired by Philips HD9
(\emph{Training} in Fig.~\ref{fig:pipeline}).
The trained networks was then fine-tuned and evaluated (\emph{Fine-tuning} and \emph{Inference} in Fig.~\ref{fig:pipeline}) on both volume-sampled 2D images and native 2D freehand images. 
The training and testing images were acquired from different clinical sites and machines,
simulating the cross-domain variance observed in reality. 
We compared \proposed with \baseline and the same fine-tuned by popular unsupervised deep DA methods, namely MK-MMD~\cite{long2015learning}, DANN~\cite{ganin2016domain} and CORAL~\cite{sun2016deep}.
Their implementation details are in the Supplementary Materials.
\\
\\
\noindent \textbf{Volume-sampled testing images.}
We tested \proposed and other baseline approaches on 2D slices sampled from 17 aligned 3D volumes acquired by GE Voluson E10,
which were different from the training volumes (Philips HD9). 
3000 slices were sampled from each testing volume.
Two evaluation metrics were used,
namely the Euclidean distance (ED) between the coordinates of the predicted and ground-truth planes in the $\mathbb{R}^{3}_{atlas}$ and the dihedral angle (DA) between them.
\\
\\
\noindent \textbf{Native 2D freehand images.}
Images from video sequences of 2D freehand ultrasound brain scans,
acquired by GE Voluson E10 and Voluson E8 from two different clinical centers, were tested and qualitatively analyzed.
As the ground-truth locations were not available,
it was not possible to achieve the same detailed quantitative analysis as the volume-sampled images.
We, thus, proposed another quantitative test.
As the acquisition of the video sequences was smooth and continuous,
the locations of consecutive images should not change abruptly,
but show a gradual transition.
We quantify such a rate of change ($\mathrm{\Delta} c$) as: 
\begin{equation}
\label{eq:deltac}
\mathrm{\Delta} c = 
\frac{\mathrm{ED}(\hat{\mathbf{P}}_i,\hat{\mathbf{P}}_{i+1})}
{1-\mathrm{NCC}(\mathbf{I}_i, \mathbf{I}_{i+1})}
\end{equation}
where $\hat{\mathbf{P}}_i$ are the coordinates of the predicted plane of $\mathbf{I}_i$ and NCC is the normalized cross-correlation. 
We used normalized (\ie by the mean of $\mathrm{\Delta} c$) standard deviation (NSTD) to quantify the consistency of $\mathrm{\Delta} c$ throughout the whole video sequence, 
which should be low ideally.
More details of the datasets are in the Supplementary Materials.

\label{sec:exp}


\section{Results and Discussion}
\label{sec:result}


\begin{table}[t]
\centering
\caption{Evaluation results (mean$\pm$standard deviation) on volume-sampled 2D images on two settings, \ref{table:same} and \ref{table:different},
evaluated by Euclidean distance (ED) and dihedral angle (DA).
The voxel size is 0.6$mm$.
$\downarrow$ indicates lower values being more accurate.
* indicates manual annotation being used.
}
\begin{subtable}[t]{.68\textwidth}
\centering
\begin{tabular}{l|l|l}
                                              & \multicolumn{1}{c|}{\textbf{\begin{tabular}[c]{@{}c@{}}ED $\downarrow$\\ (voxel)\end{tabular}}} & \multicolumn{1}{c}{\textbf{\begin{tabular}[c]{@{}c@{}}DA $\downarrow$\\ (rad)\end{tabular}}} \\ \hline
\textbf{Yeung et at.\cite{yeung2021learning}} &                                                                                                 &                                                                                              \\
\quad without fine-tuning                     & 71.1$\pm$29.9                                                                                   & 0.264$\pm$0.177                                                                              \\
\quad with MK-MMD\cite{long2015learning}      & 71.4$\pm$27.0                                                                                   & 0.266$\pm$0.153                                                                              \\
\quad with CORAL\cite{sun2016deep}            & 79.3$\pm$29.9                                                                                   & 0.276$\pm$0.159                                                                              \\
\quad with DANN\cite{ganin2016domain}         & 72.8$\pm$30.5                                                                                   & 0.265$\pm$0.160                                                                              \\
\quad *supervised fine-tuning                 & 11.3$\pm$1.57                                                                                   & 0.172$\pm$0.055                                                                              \\ \hline
\textbf{AdLocUI (ours)}                       &                                                                                                 &                                                                                              \\
\quad without fine-tuning                     & 63.0$\pm$29.0                                                                                   & 0.251$\pm$0.166                                                                              \\
\quad proposed fine-tuning          & \textbf{23.7$\pm$9.01}                                                                          & \textbf{0.198$\pm$0.092}                                                                    
\end{tabular}
\caption{Fine-tune and test on the \emph{same} set of images}
\label{table:same}
\end{subtable}
\begin{subtable}[t]{.3\textwidth}
\begin{tabular}{l|l}
\multicolumn{1}{c|}{\textbf{\begin{tabular}[c]{@{}c@{}}ED $\downarrow$\\ (voxel)\end{tabular}}} & \multicolumn{1}{c}{\textbf{\begin{tabular}[c]{@{}c@{}}DA $\downarrow$\\ (rad)\end{tabular}}} \\ \hline
                                                                                                &                                                                                              \\
70.6$\pm$25.3                                                                                   & 0.265$\pm$0.137                                                                              \\
72.6$\pm$26.0                                                                                   & 0.267$\pm$0.140                                                                              \\
80.8$\pm$28.1                                                                                   & 0.278$\pm$0.149                                                                              \\
72.4$\pm$27.9                                                                                   & 0.266$\pm$0.143                                                                              \\
28.6$\pm$14.2                                                                                   & 0.202$\pm$0.084                                                                              \\ \hline
                                                                                                &                                                                                              \\
62.7$\pm$25.0                                                                                   & 0.253$\pm$0.138                                                                              \\
\textbf{33.0$\pm$15.1}                                                                          & \textbf{0.211$\pm$0.097}                                                                    
\end{tabular}
\caption{\emph{different} set of images}
\label{table:different}
\end{subtable}
\label{table:result}
\end{table}


\subsection{Volume-Sampled Images}
\label{sec:result_synthetic}

We compared \proposed, via ablation studies, to different baseline approaches in two different settings, both corresponding to realistic scenarios.


\emph{Firstly},
as presented in Table~\ref{table:same},
we considered the scenario where the \emph{same} set of images was used for fine-tuning and then testing. 
This is relevant when \emph{offline} analysis is performed,
where we have sufficient time for fine-tuning with the test images before final analysis.
From Table~\ref{table:same}, 
the original \baseline (\ie without fine-tuning) achieved ED=71.1 and DA=0.264, 
which was slightly worse than 
\proposed without fine-tuning (ED=63.0 and DA=0.251). 
The multi-task learning (\ie additional task of predicting $\mathbf{D}_{ij}$) contributed to such improvement.
Our proposed fine-tuning step,
which does not require any additional manual annotation,
contributed to a significant (p$<$0.05, student's t-test) improvement
(ED=23.7 and DA=0.198).
We also analyzed an \emph{unlikely} situation 
where we assumed to have the ground-truth locations of the testing images
for fine-tuning (\ie retraining) \baseline in a supervised manner.
This can be viewed as the \emph{oracle} of the accuracy of the prediction (ED=11.3 and DA=0.172).

\emph{Secondly},
as presented in Table~\ref{table:different},
we considered the scenario where \emph{different} sets of images (from the same domain) were used for fine-tuning and testing. 
This corresponds to \emph{online} prediction, 
for example scanning guidance,
where a set of example images were acquired in advance for fine-tuning .
From Table~\ref{table:different},
without fine-tuning,
\baseline (ED=70.6 and DA=0.265) and \proposed (ED=62.7 and DA=0.253) performed similarly as the first scenario.
Compared to the first scenario,
a pronounced drop in performance was seen for supervised fine-tuning of \baseline (ED=28.6 and DA=0.202) when the fine-tuning and testing images were no longer the same.
This had less severe impact to \proposed with the proposed fine-tuning (ED=33.0 and DA=0.211),
which was still significantly (p$<$0.05) better than the baselines.
Despite its slightly better performance,
supervised fine-tuning requires manually annotated image locations to retrain the network for every new machine or protocol,
which is not applicable in practice.
On the contrary,
\proposed just needs the raw 2D images for fine-tuning,
which is much more achievable in neuroimaging studies.
\\
\\
\noindent \textbf{Fine-tuned with existing DA methods.}
We also compared \proposed with \baseline fine-tuned by popular unsupervised DA methods (\ie MK-MMD~\cite{long2015learning}, DANN~\cite{ganin2016domain} and CORAL~\cite{sun2016deep}).
Despite some trials of hyperparameters tuning, 
as shown in Table~\ref{table:result},
their results were still comparable or worse than no fine-tuning.
This may be due to the fact that most DA approaches were designed for classification tasks, 
which may not be directly applicable to our regression task \cite{chen2021representation}.
This further verifies the value of our work.

\subsection{Native Freehand Images}
\label{sec:result_real}
In our experiments on native 2D freehand ultrasound images,
we used the predicted image locations to sample the corresponding slices from the 3D atlas,
to which the 3D training volumes were co-aligned.
The sampled slices should match with the corresponding input images for accurate predictions.
As shown in Fig.~\ref{fig:qualitative},
predictions from \proposed clearly demonstrated a much better match, 
in terms of similarity and anatomical structures present,
with the corresponding input images at different orientations,
when compared to \baseline.
By our proposed quantitative test (\ie NSTD of $\mathrm{\Delta} c$) as described in Section \ref{sec:experiment},
\proposed achieved a result of 0.553, which was lower than
both \baseline (0.706) and \proposed without fine-tuning (0.726),
suggesting that the predicted localization of \proposed was more consistent throughout the ultrasound video sequence,
which was indicative of the smooth frame-to-frame transitions expected in freehand scanning.
Both the qualitative and quantitative results showed \proposed's superior performance when being applied on native 2D freehand ultrasound images in practice.
More qualitative examples are in the Supplementary Materials.



\section{Conclusion}
In summary, 
we propose \proposed,
a framework for localizing 2D ultrasound brain images in the 3D anatomy.
By using an intuitive cycle consistency loss,
\proposed can be fine-tuned in an \textbf{unsupervised} manner to adapt to images acquired from different machines and protocols.
The experiments on three different datasets of ultrasound images
demonstrate \proposed's generalizability and superior performance to other baseline approaches.
As future studies, we would like to extend \proposed to other anatomies and develop it as an accessible and general \emph{sensorless} freehand ultrasound guidance tool for training novice sonographers to facilitate more contextualized structural analysis and diagnosis.
\\

\par{\noindent \textbf{Acknowledgments.}} 
PH. Yeung is grateful for support from the RC Lee Centenary Scholarship. 
W. Xie is supported by the EPSRC Programme Grant Visual AI (EP/T028572/1).  
A. Namburete is funded by the UK Royal Academy of Engineering under its Engineering for Development Research Fellowship scheme and the Academy of Medical Sciences.
We thank Linde Hesse, Madeleine Wyburd and Nicola Dinsdale for their valuable suggestions and comments about the work.
For the purpose of Open Access, the author has applied a CC BY public copyright licence to any Author Accepted Manuscript version arising from this submission.

\bibliographystyle{splncs04}
\bibliography{ref_new}

\begin{thebibliography}{10}
\providecommand{\url}[1]{\texttt{#1}}
\providecommand{\urlprefix}{URL }
\providecommand{\doi}[1]{https://doi.org/#1}

\bibitem{Baumgartner2017}
Baumgartner, C.F., Kamnitsas, K., Matthew, J., Fletcher, T.P., Smith, S., Koch,
  L.M., Kainz, B., Rueckert, D.: {SonoNet}: Real-time detection and
  localisation of fetal standard scan planes in freehand ultrasound. IEEE
  Transactions on Medical Imaging  \textbf{36}(11),  2204--2215 (2017)

\bibitem{chen2020simple}
Chen, T., Kornblith, S., Norouzi, M., Hinton, G.: A simple framework for
  contrastive learning of visual representations. In: International conference
  on machine learning. pp. 1597--1607. PMLR (2020)

\bibitem{chen2021representation}
Chen, X., Wang, S., Wang, J., Long, M.: Representation subspace distance for
  domain adaptation regression. In: International Conference on Machine
  Learning. pp. 1749--1759. PMLR (2021)

\bibitem{delaunay2020unsupervised}
Delaunay, R., Hu, Y., Vercauteren, T.: An unsupervised approach to ultrasound
  elastography with end-to-end strain regularisation. In: International
  Conference on Medical Image Computing and Computer-Assisted Intervention. pp.
  573--582. Springer (2020)

\bibitem{dou2019agent}
Dou, H., Yang, X., Qian, J., Xue, W., Qin, H., Wang, X., Yu, L., Wang, S.,
  Xiong, Y., Heng, P.A., et~al.: Agent with warm start and active termination
  for plane localization in 3d ultrasound. In: International Conference on
  Medical Image Computing and Computer-Assisted Intervention. pp. 290--298.
  Springer (2019)

\bibitem{droste2020automatic}
Droste, R., Drukker, L., Papageorghiou, A.T., Noble, J.A.: Automatic probe
  movement guidance for freehand obstetric ultrasound. In: International
  Conference on Medical Image Computing and Computer-Assisted Intervention. pp.
  583--592. Springer (2020)

\bibitem{ganin2016domain}
Ganin, Y., Ustinova, E., Ajakan, H., Germain, P., Larochelle, H., Laviolette,
  F., Marchand, M., Lempitsky, V.: Domain-adversarial training of neural
  networks. The journal of machine learning research  \textbf{17}(1),
  2096--2030 (2016)

\bibitem{gao2020label}
Gao, Y., Beriwal, S., Craik, R., Papageorghiou, A.T., Noble, J.A.: Label
  efficient localization of fetal brain biometry planes in ultrasound through
  metric learning. In: Medical Ultrasound, and Preterm, Perinatal and
  Paediatric Image Analysis, pp. 126--135. Springer (2020)

\bibitem{Gao2017}
Gao, Y., Noble, J.A.: Detection and characterization of the fetal heartbeat in
  free-hand ultrasound sweeps with weakly-supervised two-streams convolutional
  networks. In: International Conference on Medical Image Computing and
  Computer-Assisted Intervention. pp. 305--313. Springer (2017)

\bibitem{Goncalves2005}
Gonçalves, L.F., Lee, W., Espinoza, J., Romero, R.: Three‐ and
  4‐dimensional ultrasound in obstetric practice: Does it help? Journal of
  Ultrasound in Medicine  \textbf{24}(12),  1599--1624 (2005)

\bibitem{grill2020bootstrap}
Grill, J.B., Strub, F., Altch{\'e}, F., Tallec, C., Richemond, P., Buchatskaya,
  E., Doersch, C., Avila~Pires, B., Guo, Z., Gheshlaghi~Azar, M., et~al.:
  Bootstrap your own latent-a new approach to self-supervised learning.
  Advances in Neural Information Processing Systems  \textbf{33},  21271--21284
  (2020)

\bibitem{hou2017predicting}
Hou, B., Alansary, A., McDonagh, S., Davidson, A., Rutherford, M., Hajnal,
  J.V., Rueckert, D., Glocker, B., Kainz, B.: Predicting slice-to-volume
  transformation in presence of arbitrary subject motion. In: International
  Conference on Medical Image Computing and Computer-Assisted Intervention. pp.
  296--304. Springer (2017)

\bibitem{hou20183}
Hou, B., Khanal, B., Alansary, A., McDonagh, S., Davidson, A., Rutherford, M.,
  Hajnal, J.V., Rueckert, D., Glocker, B., Kainz, B.: 3-d reconstruction in
  canonical co-ordinate space from arbitrarily oriented 2-d images. IEEE
  transactions on medical imaging  \textbf{37}(8),  1737--1750 (2018)

\bibitem{kz2020semi}
KZ~Tehrani, A., Mirzaei, M., Rivaz, H.: Semi-supervised training of optical
  flow convolutional neural networks in ultrasound elastography. In:
  International Conference on Medical Image Computing and Computer-Assisted
  Intervention. pp. 504--513. Springer (2020)

\bibitem{li2021autonomous}
Li, K., Wang, J., Xu, Y., Qin, H., Liu, D., Liu, L., Meng, M.Q.H.: Autonomous
  navigation of an ultrasound probe towards standard scan planes with deep
  reinforcement learning. In: 2021 IEEE International Conference on Robotics
  and Automation. pp. 8302--8308. IEEE (2021)

\bibitem{long2015learning}
Long, M., Cao, Y., Wang, J., Jordan, M.: Learning transferable features with
  deep adaptation networks. In: International conference on machine learning.
  pp. 97--105. PMLR (2015)

\bibitem{mohamed2019survey}
Mohamed, F., Siang, C.V.: A survey on 3d ultrasound reconstruction techniques.
  Artificial Intelligence—Applications in Medicine and Biology pp. 73--92
  (2019)

\bibitem{mozaffari2017freehand}
Mozaffari, M.H., Lee, W.S.: Freehand 3-d ultrasound imaging: a systematic
  review. Ultrasound in medicine \& biology  \textbf{43}(10),  2099--2124
  (2017)

\bibitem{namburete2018fully}
Namburete, A.I., Xie, W., Yaqub, M., Zisserman, A., Noble, J.A.:
  Fully-automated alignment of 3d fetal brain ultrasound to a canonical
  reference space using multi-task learning. Medical image analysis
  \textbf{46},  1--14 (2018)

\bibitem{oord2018representation}
Oord, A.v.d., Li, Y., Vinyals, O.: Representation learning with contrastive
  predictive coding. arXiv preprint arXiv:1807.03748  (2018)

\bibitem{Paladini2007}
Paladini, D., Malinger, G., Monteagudo, A., Pilu, G., Timor-Tritsch, I., Toi,
  A.: Sonographic examination of the fetal central nervous system: guidelines
  for performing the 'basic examination' and the 'fetal neurosonogram'.
  Ultrasound in Obstetrics and Gynecology  \textbf{29}(1),  109--116 (2007)

\bibitem{Papageorghiou2014}
Papageorghiou, A.T., Ohuma, E.O., Altman, D.G., Todros, T., Ismail, L.C.,
  Lambert, A., Jaffer, Y.A., Bertino, E., Gravett, M.G., Purwar, M.:
  International standards for fetal growth based on serial ultrasound
  measurements: the fetal growth longitudinal study of the {INTERGROWTH-21st}
  project. The Lancet  \textbf{384}(9946),  869--879 (2014)

\bibitem{sun2016deep}
Sun, B., Saenko, K.: Deep coral: Correlation alignment for deep domain
  adaptation. In: European conference on computer vision. pp. 443--450.
  Springer (2016)

\bibitem{yeung2021learning}
Yeung, P.H., Aliasi, M., Papageorghiou, A.T., Haak, M., Xie, W., Namburete,
  A.I.: Learning to map 2d ultrasound images into 3d space with minimal human
  annotation. Medical Image Analysis  \textbf{70},  101998 (2021)

\bibitem{yeung2021implicitvol}
Yeung, P.H., Hesse, L., Aliasi, M., Haak, M., Xie, W., Namburete, A.I., et~al.:
  Implicitvol: Sensorless 3d ultrasound reconstruction with deep implicit
  representation. arXiv preprint arXiv:2109.12108  (2021)

\bibitem{zhou2016learning}
Zhou, T., Krahenbuhl, P., Aubry, M., Huang, Q., Efros, A.A.: Learning dense
  correspondence via 3d-guided cycle consistency. In: Proceedings of the IEEE
  Conference on Computer Vision and Pattern Recognition. pp. 117--126 (2016)

\end{thebibliography}

\newpage
\section{Supplementary Materials}
\label{sec:supp}

\begin{table}[]
\centering
\caption{Implementation details of different approaches}
\begin{tabular}{|
>{\columncolor[HTML]{EFEFEF}}l |ll|}
\hline
\textit{Approaches}                                                                                                    & \multicolumn{1}{l|}{\textbf{\baseline}}                                                                                                                                                                                                                 & \textbf{\proposed}                                                                                \\ \hline
\textit{\begin{tabular}[c]{@{}l@{}}Encoder \\ $\psi_{enc}(\cdot;\theta_{enc})$\end{tabular}}                           & \multicolumn{2}{l|}{VGG16 Backbone}                                                                                                                                                                                                                                                                                                                         \\ \hline
\textit{\begin{tabular}[c]{@{}l@{}}Location prediction \\ $\psi_{loc}(\cdot;\theta_{loc})$\end{tabular}}               & \multicolumn{1}{l|}{Refer to the original paper \cite{yeung2021learning}}                                                                                                                                                                               & \begin{tabular}[c]{@{}l@{}}(FC - ReLU)$\times$3\\ \\ FC size from \\ 512 to 256 to 9\end{tabular} \\ \hline
\textit{\begin{tabular}[c]{@{}l@{}}Displacement prediction \\ $\psi_{disp}(\cdot;\theta_{disp})$\end{tabular}}         & \multicolumn{1}{c|}{-}                                                                                                                                                                                                                                  & \begin{tabular}[c]{@{}l@{}}(FC - ReLU)$\times$3\\ \\ FC size from \\ 512 to 256 to 9\end{tabular} \\ \hline
\cellcolor[HTML]{EFEFEF}                                                                                               & \multicolumn{1}{c|}{-}                                                                                                                                                                                                                                  & \begin{tabular}[c]{@{}l@{}}- $w_{L}$ = 1\\ - $w_{D}$ = 0.5\end{tabular}                           \\ \cline{2-3} 
\multirow{-2}{*}{\cellcolor[HTML]{EFEFEF}\textit{\begin{tabular}[c]{@{}l@{}}Training \\ hyperparameters\end{tabular}}} & \multicolumn{2}{l|}{\begin{tabular}[c]{@{}l@{}}- Batch size of 80\\ - Learning rate (lr) of 0.0001\\ - lr halved when errors plateaued\\ - Early stop when errors further plateaued\\ - ADAM optimization\end{tabular}}                                                                                                                                     \\ \hline
\textit{\begin{tabular}[c]{@{}l@{}}Fine-tuning \\ hyperparameters\end{tabular}}                                        & \multicolumn{1}{l|}{\begin{tabular}[c]{@{}l@{}}- Weight for MK-MMD\cite{long2015learning} loss = 10\\ - Weight for CORAL\cite{sun2016deep} loss = 1\\ - Weight for DANN\cite{ganin2016domain} loss = 1\\ - Weight for supervised loss = 1\end{tabular}} & - $w_{c}$ = 1                                                                                     \\ \hline
\textit{Other details}                                                                                                 & \multicolumn{2}{l|}{\begin{tabular}[c]{@{}l@{}}- Python 3.7, pytorch 1.9\\ - Nvidia GTX 1080ti, 12 GB memory\end{tabular}}                                                                                                                                                                                                                                   \\ \hline
\end{tabular}
\end{table}

\begin{sidewaystable}[t]
\caption{Details of the training and testing datasets}
\begin{tabular}{|
>{\columncolor[HTML]{EFEFEF}}l |llll|}
\hline
\textit{Purpose}                                                       & \multicolumn{1}{l|}{\textbf{Training}}                                                        & \multicolumn{3}{l|}{\textbf{Testing}}                                                                             \\ \hline
\textit{Type}                                                          & \multicolumn{1}{l|}{3D volumes}                                                               & \multicolumn{1}{l|}{3D volumes}        & \multicolumn{2}{l|}{2D freehand images}                                  \\ \hline
\textit{Number}                                                        & \multicolumn{1}{l|}{50}                                                                       & \multicolumn{1}{l|}{17}                & \multicolumn{1}{l|}{4 sequences (829 images)} & 3 sequences (531 images) \\ \hline
\textit{\begin{tabular}[c]{@{}l@{}}Acquisition\\ details\end{tabular}} & \multicolumn{1}{l|}{Philips HD9 curvilinear probe at a 2–5 MHz}                               & \multicolumn{2}{l|}{GE Voluson E10}                                                    & GE Voluson E8            \\ \hline
\textit{Dimensions}                                                    & \multicolumn{1}{l|}{$160^3$ voxels at a resolution of $0.6\times0.6\times0.6\ \mathrm{mm}^3$} & \multicolumn{1}{l|}{Resize to $160^3$} & \multicolumn{2}{l|}{Crop to $160^2$}                                     \\ \hline
\textit{Gestational age}                                               & \multicolumn{4}{l|}{19-21 gestational weeks}                                                                                                                                                                      \\ \hline
\textit{Related studies}                                               & \multicolumn{1}{l|}{the INTERGROWTH-21st study~\cite{Papageorghiou2014}}                      & \multicolumn{1}{c|}{-}                 & \multicolumn{1}{c|}{-}                        & \multicolumn{1}{c|}{-}   \\ \hline
\end{tabular}
\end{sidewaystable}

\label{sec:supp}

\end{document}